\documentclass[conference]{IEEEtran}
\IEEEoverridecommandlockouts
\usepackage{amsmath,amssymb,amsfonts}
\usepackage[ruled]{algorithm2e}
\usepackage{bm}
\usepackage{cite}
\usepackage{graphicx}
\usepackage{textcomp}
\usepackage{xcolor}
\usepackage{url}
\usepackage{siunitx}
\newtheorem{proposition}{\underline{Proposition}}

\def\BibTeX{{\rm B\kern-.05em{\sc i\kern-.025em b}\kern-.08em
		T\kern-.1667em\lower.7ex\hbox{E}\kern-.125emX}}

\begin{document}
	\title{Hybrid Beamforming Design for Integrated Sensing and Communication Exploiting Prior Information}
	\author{\IEEEauthorblockN{Yizhuo Wang and Shuowen Zhang\\
			Department of Electrical and Electronic Engineering, The Hong Kong Polytechnic University\\}
		E-mail: yizhuo-eee.wang@connect.polyu.hk, shuowen.zhang@polyu.edu.hk
		}
		
	\maketitle
	
	\begin{abstract}
		In this paper, we investigate the hybrid beamforming design for a multiple-input multiple-output (MIMO) integrated sensing and communication (ISAC) system, where a multi-antenna base station (BS) with hybrid analog-digital transmit antenna arrays sends dual-functional signals to communicate with a multi-antenna user and simultaneously sense the location information of a point target based on the reflected echo signals. Specifically, we aim to sense the target's \emph{unknown} and \emph{random} angle information by exploiting its prior distribution information, with \emph{posterior Cram\'{e}r-Rao bound (PCRB)} employed as the sensing performance metric. First, we consider a sensing-only case and study the hybrid beamforming optimization to minimize the sensing PCRB. We analytically prove that hybrid beamforming can achieve the same performance as the optimized digital beamforming as long as the number of radio frequency (RF) chains is larger than 1. Then, we propose a convex relaxation based algorithm for the hybrid beamforming design with a single RF chain. Next, we study the hybrid beamforming optimization to minimize the PCRB subject to a communication rate target. Due to the intractability of the exact PCRB expression, we replace it with a tight upper bound. Although this problem is still non-convex and challenging to solve, we propose an alternating optimization (AO) algorithm for finding a high-quality suboptimal solution based on the feasible point pursuit successive convex approximation (FPP-SCA) method. Numerical results validate the effectiveness of our proposed hybrid beamforming design.
	\end{abstract}
	
	\section{Introduction}
	Recently, integrated sensing and communication (ISAC) has attracted increasing research interests to support various new applications, e.g., industrial internet-of-things (IIoT), vehicles-to-everything (V2X), and smart home \cite{liu2022integrated}. To harness the full potential of ISAC, extensive research has been conducted on the design of transmitted signals or beamforming. In \cite{liu2020joint}, the weighted summation of independent radar waveforms and communication symbols was transmitted to increase the degrees-of-freedom (DoFs) for multiple-input multiple-output (MIMO) radar. In \cite{liu2022joint}, the transmitted signals were designed to maximize the radar output signal-to-interference-plus-noise ratio (SINR). The \emph{Cram\'{e}r-Rao bound (CRB)} is well known as a mean-squared error (MSE) lower bound, which can explicitly reflect the sensing performance. Thus, it has been widely used as a performance metric for transmitted signal design in ISAC systems \cite{liu2021cramer}, \cite{hua2024mimo}. However, CRB is only suitable for the case where the parameters to be sensed are \emph{known} and \emph{deterministic}, and the transmitted signals can only be designed based on the CRB corresponding to a given parameter (e.g., target's location). In practical scenarios, the parameters to be sensed can be \emph{unknown} and \emph{random}, for which the distributions can be known a \emph{priori} for exploitation. In this case, \emph{posterior Cram\'{e}r-Rao bound (PCRB)} can characterize the MSE lower bound with the prior information exploited \cite{van1968detection}. Along this line, \cite{xu2023radar,xu2023mimo,xu2024isac} studied the optimal transmitted signal design for a MIMO radar or a MIMO ISAC system, where useful properties of the optimal transmit covariance matrix were unveiled. Particularly, it was shown that the transmit signal design is critically dependent on the prior distribution information. Moreover, \cite{hou2023secure,hou2023optimal} studied the transmit signal optimization in a secure ISAC system where the target serves as a potential eavesdropper.
	
	\begin{figure}[t]
		\vspace{-3mm}
		\centering
		\includegraphics[width=0.4\textwidth]{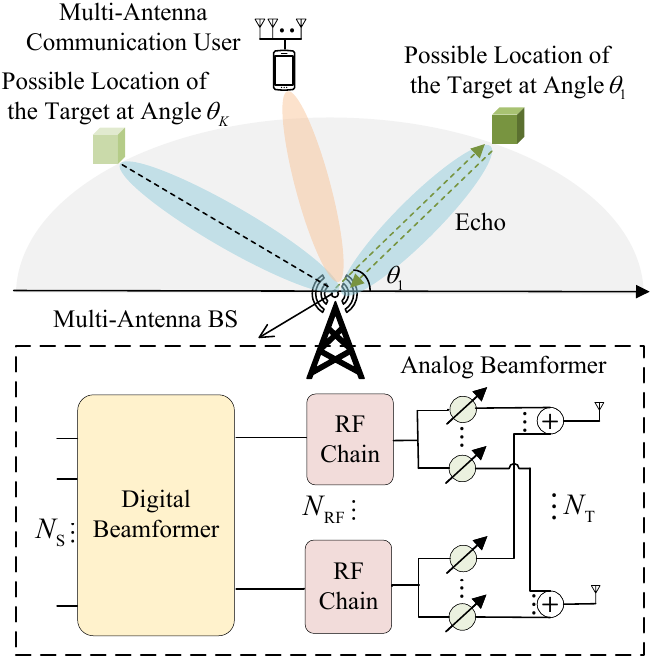}
		\vspace{-3mm}
		\caption{Illustration of a MIMO ISAC system with hybrid beamforming.}
		\label{System_Model}
		\vspace{-6mm}
	\end{figure} 
	
	The above works on ISAC exploiting prior information only focused on fully-digital transmit beamforming design. To further enhance the communication rate and sensing accuracy, millimeter-wave (mmWave) and large-scale MIMO are envisioned as two important technologies, which, however, call for the adoption of the \emph{hybrid analog-digital} architecture due to the high cost of equipping each antenna with a dedicated radio frequency (RF) chain. This thus motivates our study of \emph{hybrid beamforming} in MIMO ISAC systems exploiting prior information. It is worth noting that although there have been existing works on hybrid beamforming in communication systems \cite{el2014spatially,sohrabi2016hybrid,yu2016alternating} and ISAC systems \cite{liu2019hybrid,gong2023hybrid,wang2022partially}, how to leverage the prior information for hybrid beamforming design in sensing or ISAC systems still remains an open problem to the best of our knowledge.
	
	This paper studies the hybrid beamforming design in a MIMO ISAC system, where a multi-antenna base station (BS) equipped with hybrid analog-digital transmit array sends dual-functional signals to communicate with a multi-antenna user and sense the \emph{unknown} and \emph{random} location parameter of a point target, as illustrated in Fig. \ref{System_Model}. The prior distribution information of the location parameter is known for exploitation. We first consider the sensing-only case and formulate the hybrid beamforming optimization problem for PCRB minimization. We analytically prove that two RF chains suffice for hybrid beamforming to achieve the same optimal PCRB performance as fully-digital beamforming, and further devise an efficient hybrid beamforming algorithm with a single RF chain. Then, we study the hybrid beamforming optimization problem to minimize a tight PCRB upper bound under a communication rate target, which is a challenging non-convex problem. We propose an alternating optimization (AO) algorithm based on the feasible point pursuit successive convex optimization (FPP-SCA) method \cite{mehanna2014feasible} for finding a high-quality suboptimal solution. It is shown via numerical results that our proposed design outperforms various benchmark schemes.
	
	\addtolength{\topmargin}{0.03in}
	\section{System Model}
	We consider a narrowband MIMO ISAC system which consists of a multi-antenna BS equipped with $N_{\mathrm{T}}\geq 1$ transmit antennas and $N_{\mathrm{R}}\geq 1$ co-located receive antennas, a multi-antenna user equipped with $N_{\mathrm{U}}\geq 1$ antennas, and a point target whose \emph{unknown} and \emph{random} location information needs to be sensed. We consider downlink transmission where the BS sends dual-functional signals to communicate with the user and estimate the location information of the target via the echo signals reflected by the target and received back at the BS receive antennas. We focus on a challenging case where the BS transmitter adopts a fully-connected \emph{hybrid analog-digital} array structure with only $N_{\mathrm{RF}}<N_{\mathrm{T}}$ RF chains, as shown in Fig. \ref{System_Model}. Note that this is a practical case in various scenarios, e.g., when the system operates over mmWave frequency bands or when the number of BS transmit antennas is large for achieving high sensing accuracy and communication rate.
	
	We assume that the BS-target distance (range) is known \emph{a priori} as $r$ in meters (m), e.g., via time-of-arrival (ToA) methods. Therefore, we only need to focus on the estimation of the target's \emph{angle} $\theta$ with respect to the BS, as illustrated in Fig. \ref{System_Model}. The probability density function (PDF) of $\theta$ is denoted by $p_{\Theta}(\theta)$, which is assumed to be known \emph{a priori} based on target movement pattern or statistical information. We consider a half-wavelength spaced uniform linear array (ULA) layout at the BS, and the angle to be sensed lies in the range $[-\frac{\pi}{2}, \frac{\pi}{2})$.
	
	Let $N_{\mathrm{S}}$ denote the number of data streams with $N_{\mathrm{S}}\leq N_{\mathrm{RF}}$, which is also a design variable. Suppose that $L\geq 1$ symbol intervals are used to estimate $\theta$ and $\bm{s}_l\in \mathbb{C}^{N_{\mathrm{S}}\times 1}$ denotes the data symbol vector at the $l$-th interval. The collection of $\bm{s}_l$ over $L$ symbol intervals is denoted by $\bm{S}=[\bm{s}_1,\cdots,\bm{s}_L]$ with $\frac{1}{L}\bm{S}\bm{S}^H=\bm{I}_{N_{\mathrm{S}}}$. The baseband equivalent transmitted signal vector at the $l$-th interval is given by
	\begin{equation}
		\bm{x}_l=\bm{F}_{\mathrm{RF}}\bm{F}_{\mathrm{BB}}\bm{s}_l,\quad l=1,\cdots,L,
	\end{equation}
	where $\bm{F}_{\mathrm{RF}}\in \mathbb{C}^{N_{\mathrm{T}}\times N_{\mathrm{RF}}}$ denotes the analog beamformer realized by phase shifters, $\bm{F}_{\mathrm{BB}}\in \mathbb{C}^{N_{\mathrm{RF}}\times N_{\mathrm{S}}}$ denotes the digital beamformer. Note that every element in the analog beamformer $\bm{F}_{\mathrm{RF}}$ should satisfy the unit-modulus constraint, i.e., $|[\bm{F}_{\mathrm{RF}}]_{m,n}|=1, \forall m,n$. 
	The transmit (sample) covariance matrix is thus given by
	\begin{equation}
		\bm{R}_X\!=\!\mathbb{E}[\bm{x}_l\bm{x}_l^H]\!=\!\bm{F}_{\mathrm{RF}}\bm{F}_{\mathrm{BB}}\bm{F}_{\mathrm{BB}}^H\bm{F}_{\mathrm{RF}}^H.
	\end{equation}
	Let $P$ denote the transmit power budget, which yields $\mathrm{tr}(\bm{R}_X)\leq P$. We consider a quasi-static block fading channel between the BS and the communication user. The channel matrix in the $L$ symbol intervals of interest is assumed to remain constant and denoted by $\bm{H}\in \mathbb{C}^{N_{\mathrm{U}}\times N_{\mathrm{T}}}$, which is perfectly known at the BS and the user. The received signal vector at the communication user in each $l$-th interval is thus given by
	\begin{equation}
		\bm{y}_l^{\mathrm{C}}=\bm{H}\bm{F}_{\mathrm{RF}}\bm{F}_{\mathrm{BB}}\bm{s}_l+\bm{n}_l^{\mathrm{C}},  \quad l=1,\cdots,L,
	\end{equation}
	where \textcolor{black}{$\bm{n}_l^{\mathrm{C}}\sim\mathcal{CN}(\bm{0}, \sigma_{\mathrm{C}}^2\bm{I}_{N_{\mathrm{U}}})$} denotes the circularly symmetric complex Gaussian (CSCG) noise vector at the user receiver, with \textcolor{black}{$\sigma_{\mathrm{C}}^2$} denoting the average noise power. The achievable rate for the communication user can be expressed as
	\begin{equation}\label{rate}
		R=\log_2\bigg|\bm{I}_{N_{\mathrm{U}}}+\frac{\bm{H}\bm{F}_{\mathrm{RF}}\bm{R}_{\mathrm{BB}}\bm{F}_{\mathrm{RF}}^H\bm{H}^H}{\textcolor{black}{\sigma_{\mathrm{C}}^2}}\bigg|
	\end{equation}
	in bits per second per Hertz (bps/Hz).
	
	By sending $\bm{x}_l$, the BS can also estimate $\theta$ based on the received echo signal reflected by the target. We consider a line-of-sight (LoS) channel between the BS and the target. The target response matrix is given by
	\begin{equation}\label{G}
		\bm{G}(\theta)=\alpha\bm{b}(\theta)\bm{a}^H(\theta).
	\end{equation}
	In (\ref{G}), $\alpha\triangleq \frac{\beta_0}{r^2}\psi=\alpha_{\mathrm{R}}+j\alpha_{\mathrm{I}}\in \mathbb{C}$ denotes the reflection coefficient, which contains both the round-trip channel power $\frac{\beta_0}{r^2}$ with $\beta_0$ denoting the reference channel power at distance $1$ m, and the radar cross-section (RCS) coefficient, $\psi\in\mathbb{C}$. Note that $\alpha$ is an \emph{unknown deterministic} parameter. Moreover, $\bm{a}(\theta)\in \mathbb{C}^{N_{\mathrm{T}}\times 1}$ and $\bm{b}(\theta)\in \mathbb{C}^{N_{\mathrm{R}}\times 1}$ are steering vectors of the transmit and receive antennas, respectively. Each element in $\bm{a}(\theta)$ and $\bm{b}(\theta)$ is given by $a_p(\theta)=e^{\frac{-j\pi (N_{\mathrm{T}}-2p+1)\sin\theta}{2}}, p=1,\cdots,N_{\mathrm{T}}$, $b_q(\theta)=e^{\frac{-j\pi (N_{\mathrm{R}}-2q+1)\sin\theta}{2}}, q=1,\cdots,N_{\mathrm{R}}$. The reflected echo signal vector at the $l$-th symbol interval is given by
	\begin{equation}
		\bm{y}_l^{\mathrm{S}}=\bm{G}(\theta)\bm{F}_{\mathrm{RF}}\bm{F}_{\mathrm{BB}}\bm{s}_l+\bm{n}_l^{\mathrm{S}},  \quad l=1,\cdots,L,
	\end{equation}
	where \textcolor{black}{$\bm{n}_l^{\mathrm{S}}\sim\mathcal{CN}(\bm{0}, \sigma_{\mathrm{S}}^2\bm{I}_{N_{\mathrm{R}}})$} denotes the CSCG noise vector at the BS receiver. Therefore, the reflected echo signal matrix at the receiver is given by
	\begin{equation}
		\bm{Y}=[\bm{y}_1^{\mathrm{S}},\cdots,\bm{y}_L^{\mathrm{S}}]=\bm{G}(\theta)\bm{F}_{\mathrm{RF}}\bm{F}_{\mathrm{BB}}\bm{S}+[\bm{n}_1^{\mathrm{S}},\cdots,\bm{n}_L^{\mathrm{S}}].
	\end{equation}
	
	We employ the PCRB as the sensing performance metric, which quantifies a lower bound of the sensing MSE with prior distribution information. Note that besides $\theta$, $\alpha_{\mathrm{R}}$ and $\alpha_{\mathrm{I}}$ are also unknown parameters which need to be jointly estimated with $\theta$. Define $\bm{\zeta}=[\theta,\alpha_{\mathrm{R}},\alpha_{\mathrm{I}}]^T$ as the collection of all unknown parameters. According to \cite{xu2023radar}, the PCRB for estimating the target's angle $\theta$ is given by \textcolor{black}{
	\begin{align}\label{PCRB_expression}
		\mathrm{PCRB}_{\theta}
		&=\frac{\sigma_{\mathrm{S}}^2}{2|\alpha|^2L}\Bigg/\Bigg(\frac{\sigma_{\mathrm{S}}^2}{2|\alpha|^2L}\int \bigg(\frac{\partial\ln \big(p_{\Theta}(\theta)\big)}{\partial\theta}\bigg)^2p_{\Theta}(\theta)d\theta\nonumber\\
		&+\mathrm{tr}(\bm{A}_1\bm{R}_X)+\mathrm{tr}(\bm{A}_2\bm{R}_X)-\frac{|\mathrm{tr}(\bm{A}_3\bm{R}_X|^2}{\mathrm{tr}(\bm{A}_4\bm{R}_X)}
		\Bigg), 
	\end{align}
}
	where $\bm{A}_1=\int \|\Dot{\bm{b}}(\theta)\|^2\bm{a}(\theta)\bm{a}^H(\theta) p_{\Theta}(\theta)d\theta$, $\bm{A}_2=N_{\mathrm{R}}\int \Dot{\bm{a}}(\theta)\Dot{\bm{a}}^H(\theta) p_{\Theta}(\theta)d\theta$, $\bm{A}_3=N_{\mathrm{R}}\int \Dot{\bm{a}}(\theta)\bm{a}^H(\theta) p_{\Theta}(\theta)d\theta$, and $\bm{A}_4=N_{\mathrm{R}}\int \bm{a}(\theta)\bm{a}^H(\theta)p_{\Theta}(\theta)d\theta$, with $\Dot{\bm{x}}(\theta)$ denoting the derivative of $\bm{x}(\theta)$ with respect to $\theta$.
	
	Note that both the communication rate $R$ in (\ref{rate}) and $\mathrm{PCRB}_\theta$ in (\ref{PCRB_expression}) are functions of the transmit covariance matrix $\bm{R}_X=\bm{F}_{\mathrm{RF}}\bm{F}_{\mathrm{BB}}\bm{F}_{\mathrm{BB}}^H\bm{F}_{\mathrm{RF}}^H$ and equivalently the analog and digital beamformers $\bm{F}_{\mathrm{RF}}$ and $\bm{F}_{\mathrm{BB}}$. For ease of exposition, we define $\bm{R}_{\mathrm{BB}}\overset{\Delta}{=}\bm{F}_{\mathrm{BB}}\bm{F}_{\mathrm{BB}}^H\succeq \bm{0}$. Note that optimizing $\bm{R}_{\mathrm{BB}}$ is equivalent to optimizing $\bm{F}_{\mathrm{BB}}$ and the number of data streams $N_{\mathrm{S}}$. In the following, we first study the joint optimization of $\bm{F}_{\mathrm{RF}}$ and $\bm{R}_{\mathrm{BB}}$ to minimize the sensing PCRB without considering the communication performance, to draw fundamental insights on the optimal hybrid beamforming design for sensing exploiting prior information; then, we jointly optimize $\bm{F}_{\mathrm{RF}}$ and $\bm{R}_{\mathrm{BB}}$ to achieve an optimal trade-off between the sensing PCRB and the communication rate.
	\section{Hybrid Beamforming Optimization for PCRB Minimization}
	\subsection{Problem Formulation}
	In this section, we focus on the sensing performance and aim to optimize the hybrid beamforming design to minimize the PCRB of the MSE in sensing $\theta$ subject to a total power budget $P$. The problem is formulated as
	\begin{subequations}\label{P1}
		\begin{align}
			\mbox{(P1)}\quad \underset{\bm{R}_{\mathrm{BB}},\bm{F}_\mathrm{RF}}{\min}& \mathrm{PCRB}_{\theta} \label{P1_obj}\\
			\mbox{s.t.}\quad & \mathrm{tr}(\bm{F}_{\mathrm{RF}}\bm{R}_{\mathrm{BB}}\bm{F}_{\mathrm{RF}}^H)\leq P,\label{P1_power constraint}\\
			& |[\bm{F}_{\mathrm{RF}}]_{m,n}|=1,\quad \forall m,n \label{P1_unit-modulus constraint}\\
			& \bm{R}_{\mathrm{BB}} \succeq \bm{0}.\label{P1_semi constraint}
		\end{align}
	\end{subequations}
	\subsection{Proposed Solution to (P1)}
	Problem (P1) is a non-convex optimization problem due to the unit-modulus constraints in (9c). Nevertheless, by exploiting its structure, we have the following proposition.
	\begin{proposition}\label{prop_solution}
	When $N_{\mathrm{RF}}\geq 2$, hybrid beamforming can achieve the same PCRB performance as the optimal fully-digital beamforming.
	\end{proposition}
	\begin{IEEEproof}
	With fully-digital beamforming, there exists a rank-one optimal solution of the transmit covariance matrix $\bm{R}_X=\bm{F}_{\mathrm{RF}}\bm{R}_{\mathrm{BB}}\bm{F}_{\mathrm{RF}}^H$ to the same PCRB minimization problem, i.e., (P1) without the constraints in (\ref{P1_unit-modulus constraint}) \cite{xu2023mimo}, which implies that the optimal number of data streams for sensing is $N_{\mathrm{S}}=1$. According to \cite{zhang2005phase,sohrabi2016hybrid}, hybrid beamforming can realize fully-digital beamforming if and only if $N_{\mathrm{RF}}\geq 2N_{\mathrm{S}}=2$.
	\end{IEEEproof}

	Thus, if $N_{\mathrm{RF}}\geq2$, we can obtain the optimal fully-digital beamforming design according to \cite{xu2023mimo}, and then construct the optimal hybrid beamforming design with the same performance according to the method in \cite{sohrabi2016hybrid}, which is the optimal solution to (P1). In the following, we study (P1) for the case with insufficient RF chains, i.e., $N_{\mathrm{RF}}=1$. 
	
	 With $N_{\mathrm{RF}}=1$, $\bm{R}_{\mathrm{BB}}\in \mathbb{C}^{N_{\mathrm{RF}}\times N_{\mathrm{RF}}}$ reduces to a scalar and the optimal solution is given by ${R}_{\mathrm{BB}}^{\star}=\frac{P}{N_{\mathrm{T}}}$. The optimal digital beamformer can be written as ${F}_{\mathrm{BB}}^{\star}=\sqrt{\frac{P}{N_{\mathrm{T}}}}$. Thus, our focus is the optimization of the analog beamformer, for which we propose a convex relaxation based AO algorithm. Specifically, $\bm{F}_{\mathrm{RF}}$ reduces to a vector $\mathbf{f}_{\mathrm{RF}}\in \mathbb{C}^{N_{\mathrm{T}}\times 1}$. The PCRB minimization problem for each $m$-th element $f_m$ in $\mathbf{f}_{\mathrm{RF}}$ with all the other elements being fixed can be expressed as
	\vspace{-2mm}
		\begin{equation}\label{P1I_obj}
			\mbox{(P1-m)}\, \underset{f_m:|f_m|=1}{\max}\quad 
			g_1(f_m)+g_2(f_m)-\frac{g_3(f_m)g_3^*(f_m)}{g_4(f_m)},
		\end{equation}
	where $g_i(f_m)=\alpha_i f_m+\alpha^*_i f_m^*+\rho_i,  i=1,2,4$, $g_3(f_m)=\alpha_3 f_m+\beta_3 f_m^*+\rho_3$, $\alpha_i=\sum_{n\neq m}^{N_{\mathrm{T}}}f_n^*[\bm{A}_i]_{n,m}, i=1,2,3,4$, $\beta_3=\sum_{n\neq m}^{N_{\mathrm{T}}}f_n[\bm{A}_3]_{m,n}$, $\rho_i=\sum_{n\neq m}^{N_{\mathrm{T}}}\sum_{\substack{k\neq n,\\k\neq m}}^{N_{\mathrm{T}}}\!\! f_n^*[\bm{A}_i]_{n,k}f_k+\mathrm{tr}(\bm{A}_i),  i=1,2,3,4$, 
	with $[\bm{A}_i]_{n,k}$ denoting the $(n,k)$-th element in $\bm{A}_i$. Note that by relaxing the unit-modulus constraint to a convex constraint $|f_m|\leq 1$, (P1-m) becomes a convex optimization problem whose optimal solution can be obtained via CVX. Thus, we propose to iteratively solve (P1-m) under the relaxed constraint for $m=1,\dots,N_{\mathrm{T}}$ until convergence, which is guaranteed since the optimal solution is obtained in each iteration. Finally, each $f_m$ is normalized such that $|f_m|=1$ holds if the obtained $f_m$ does not satisfy the unit-modulus constraint.
	
	To summarize, the optimal solution to (P1) can be obtained based on the optimal digital beamformer when $N_{\mathrm{RF}}\geq 2$, and a suboptimal solution can be obtained via the convex relaxation based AO algorithm when $N_{\mathrm{RF}}=1$.
	\vspace{-1mm}		
	\addtolength{\topmargin}{0.01in}
\section{Hybrid Beamforming Optimization for PCRB-Rate Trade-Off} \label{section4}
	\vspace{-1.5mm}	
	\subsection{Problem Formulation}
	In this section, we aim to optimize the hybrid beamforming design to achieve the optimal trade-off between the sensing PCRB and the communication rate. Note that in this case, the optimal number of data streams may not be $1$, thus the results in Section III are generally not applicable. Moreover, the exact $\mathrm{PCRB}_{\theta}$ in (\ref{PCRB_expression}) has a complicated fractional form for general-rank $\bm{R}_X$ (in contrast to rank-one $\bm{R}_X$ with $N_{\mathrm{RF}}=1$ in Section III). The matrices $\bm{A}_1$, $\bm{A}_2$, $\bm{A}_3$, and $\bm{A}_4$ in (\ref{PCRB_expression}) involve complex integrals which are difficult to compute especially when $N_{\mathrm{T}}$ is very large. Furthermore, $\bm{R}_X$ is a quadratic function of $\bm{F}_{\mathrm{RF}}$ which is also coupled with $\bm{R}_{\mathrm{BB}}$. To overcome the above challenges, we propose to adopt a tight upper bound of the exact $\mathrm{PCRB}_{\theta}$ as the sensing performance metric in this section, which is given by \cite{xu2023radar}  
	\vspace{-1.5mm} \textcolor{black}{
	\begin{align}\label{PCRB upper bound}
		&\mathrm{PCRB}_{\theta}\leq\mathrm{PCRB}_{\theta}^{\mathrm{U}} \nonumber\\
		&\triangleq\frac{\frac{\sigma_{\mathrm{S}}^2}{2|\alpha|^2L}}{\frac{\sigma_{\mathrm{S}}^2}{2|\alpha|^2L}\int \big(\frac{\partial\ln (p_{\Theta}(\theta))}{\partial\theta}\big)^2p_{\Theta}(\theta)d\theta+\mathrm{tr}(\bm{A}_1\bm{R}_X)}.
	\end{align}
}
	We aim to jointly optimize $\bm{F}_{\mathrm{RF}}$ and $\bm{R}_{\mathrm{BB}}$ to minimize the PCRB upper bound, subject to a communication rate target denoted by $\Bar{R}$ bps/Hz and a total transmit power budget $P$. The problem is formulated as 
	\begin{subequations}\label{P2}
		\begin{align}
			\hspace{-4mm}
			\mbox{(P2)}\, \underset{\bm{R}_{\mathrm{BB}},\bm{F}_\mathrm{RF}}{\max}& \mathrm{tr}\big(\bm{F}_{\mathrm{RF}}\bm{R}_{\mathrm{BB}}\bm{F}_{\mathrm{RF}}^H\bm{A}_1\big) \label{P2_obj}\\
			\mbox{s.t.}\quad & \textcolor{black}{\log_2\bigg|\bm{I}_{N_{\mathrm{U}}}+\frac{\bm{H}\bm{F}_{\mathrm{RF}}\bm{R}_{\mathrm{BB}}\bm{F}_{\mathrm{RF}}^H\bm{H}^H}{\sigma_{\mathrm{C}}^2}\bigg|\geq \Bar{R}} \label{P2_rate constraint}\\
			& \mathrm{tr}(\bm{F}_{\mathrm{RF}}\bm{R}_{\mathrm{BB}}\bm{F}_{\mathrm{RF}}^H)\leq P \label{P2_power constraint}\\
			& |[\bm{F}_{\mathrm{RF}}]_{m,n}|=1, \quad \forall m,n\label{P2_unit-modulus constraint}\\
			& \bm{R}_{\mathrm{BB}}\succeq \bm{0} \label{P2_semi constraint}.
		\end{align}
	\end{subequations}
	Note that (P2) is a non-convex problem due to the unit-modulus constraints in (\ref{P2_unit-modulus constraint}). In addition, since $\bm{F}_{\mathrm{RF}}$ and $\bm{R}_{\mathrm{BB}}$ are coupled with each other in the objective function as well as (\ref{P2_rate constraint}) and (\ref{P2_power constraint}), the optimal solution to (P2) is particularly difficult to obtain. In the following, we propose an AO algorithm for obtaining a high-quality suboptimal solution to (P2), where the digital beamformer and the analog beamformer are iteratively designed with the other one being fixed at each time.
	\vspace{-2.5mm}
	\subsection{Proposed Solution to (P2)}
	\vspace{-2mm}
	\subsubsection{Digital Beamformer Optimization}
	First, we optimize the covariance matrix of the digital beamformer under a given $\bm{F}_{\mathrm{RF}}$. Let $\Tilde{\bm{H}}=\bm{H}\bm{F}_{\mathrm{RF}}$ denote the effective channel. The digital beamformer optimization problem is formulated as
	\begin{subequations}\label{P3}
		\begin{align}
			\mbox{(P3)}\quad  \underset{\bm{R}_{\mathrm{BB}}}{\max}\ & \mathrm{tr}\big(\bm{F}_{\mathrm{RF}}\bm{R}_{\mathrm{BB}}\bm{F}_{\mathrm{RF}}^H\bm{A}_1\big) \label{P3_obj}\\
			\mbox{s.t.}\quad & \textcolor{black}{\log_2\bigg|\bm{I}_{N_{\mathrm{U}}}+\frac{\Tilde{\bm{H}}\bm{R}_{\mathrm{BB}}\Tilde{\bm{H}}^H}{\sigma_{\mathrm{C}}^2}\bigg|\geq \Bar{R} } \label{P3_rate constraint}\\
			&\mathrm{tr}(\bm{F}_{\mathrm{RF}}\bm{R}_{\mathrm{BB}}\bm{F}_{\mathrm{RF}}^H)\leq P \label{P3_powerconstraint}\\
			& \bm{R}_{\mathrm{BB}} \succeq \bm{0}.
		\end{align}
	\end{subequations}
	Note that (P3) is a convex optimization problem. The optimal solution to (P3) denoted by $\bm{R}_{\mathrm{BB}}^{\star}$ can be obtained via the Lagrange duality based method in \cite{xu2023mimo}. The optimal digital beamformer $\bm{F}_{\mathrm{BB}}^{\star}$ can be obtained via eigenvalue decomposition (EVD) of $\bm{R}_{\mathrm{BB}}^{\star}$, with $N_{\mathrm{S}}^\star=\mathrm{rank}(\bm{R}_{\mathrm{BB}}^\star)$. 
	\subsubsection{Analog Beamformer Optimization}
	Next, we optimize the analog beamformer $\bm{F}_\mathrm{RF}$ under a given $\bm{R}_{\mathrm{BB}}$. The analog beamformer optimization problem is formulated as
	\begin{subequations}\label{P4}
		\begin{align}
			\mbox{(P4)}\quad  \underset{\bm{F}_\mathrm{RF}}{\max}\ & \mathrm{tr}\big(\bm{F}_{\mathrm{RF}}\bm{R}_{\mathrm{BB}}\bm{F}_{\mathrm{RF}}^H\bm{A}_1\big) \label{P4_obj}\\
			\mbox{s.t.}\quad & \textcolor{black}{\log_2\bigg|\bm{I}_{N_{\mathrm{U}}}+\frac{\bm{H}\bm{F}_{\mathrm{RF}}\bm{R}_{\mathrm{BB}}\bm{F}_{\mathrm{RF}}^H\bm{H}^H}{\sigma_{\mathrm{C}}^2}\bigg|\geq \Bar{R} }\label{P4_rate constraint}\\
			& \mathrm{tr}(\bm{F}_{\mathrm{RF}}\bm{R}_{\mathrm{BB}}\bm{F}_{\mathrm{RF}}^H)\leq P \label{P4_power constraint}\\
			& |[\bm{F}_{\mathrm{RF}}]_{m,n}|=1, \forall m,n\label{P4_unit-modulus constraint},
		\end{align}
	\end{subequations}
	which is still a non-convex problem due to the unit-modulus constraints in (\ref{P4_unit-modulus constraint}). Furthermore, the left-hand side of the rate constraint in (\ref{P4_rate constraint}) is not a concave function with respect to $\bm{F}_{\mathrm{RF}}$. Inspired by the weighted minimum mean-squared error (WMMSE) method \cite{shi2011iteratively}, we propose to convert the rate constraint in (\ref{P4_rate constraint}) into a tractable form as follows. 
	
	We first introduce a decoding matrix $\bm{Q}^H\in \mathbb{C}^{N_{\mathrm{S}}\times N_{\mathrm{U}}}$. The decoded signal vector at the $l$-th interval is thus given by $\Hat{\bm{s}}_l=\bm{Q}^H\bm{y}_l^{\mathrm{C}}$. The MSE matrix for information decoding is given by
	\vspace{-5mm}
	\begin{align}\label{MSE}
		&\bm{E}\!=\!\mathbb{E}\Big[(\Hat{\bm{s}}_l-\bm{s}_l)(\Hat{\bm{s}}_l-\bm{s}_l)^H\Big] \\
		&\!\!=\!\!(\!\bm{Q}^H\!\bm{H}\!\bm{F}_{\mathrm{RF}}\!\bm{F}_{\mathrm{BB}\!}\!-\!\bm{I}_{N_{\mathrm{S}}}\!)(\!\bm{Q}^H\!\bm{H}\!\bm{F}_{\mathrm{RF}}\!\bm{F}_{\mathrm{BB}}\!-\!\bm{I}_{N_{\mathrm{S}}}\!)^H\!\!+\!\!\textcolor{black}{\sigma_{\mathrm{C}}^2}\bm{Q}^H\bm{Q}.\nonumber
	\end{align}  
	The communication rate can be expressed as
	\begin{equation}\label{reformulate_rate constraint}
		\xi(\bm{W}, \bm{Q}, \bm{F}_{\mathrm{RF}})\!=\!\log_2\big|\bm{W}\big|\!-\!\mathrm{tr}(\bm{W}\bm{E})\!+\!N_{\mathrm{S}},
	\end{equation}
	where $\bm{W}\in \mathbb{C}^{N_{\mathrm{S}}\times N_{\mathrm{S}}}$ is an auxiliary matrix. 
	
	With given $\bm{F}_{\mathrm{RF}}$, the optimal $\bm{Q}$ for maximizing $\xi(\bm{W}, \bm{Q}, \bm{F}_{\mathrm{RF}})$ can be obtained based on the first-order optimality condition, which is given by $\bm{Q}^{\star}=\bm{J}^{-1}\bm{H}\bm{F}_{\mathrm{RF}}\bm{F}_{\mathrm{BB}}$,
	with \textcolor{black}{$\bm{J}=\sigma_{\mathrm{C}}^2\bm{I}_{N_{\mathrm{U}}}+\bm{H}\bm{F}_{\mathrm{RF}}\bm{F}_{\mathrm{BB}}\bm{F}_{\mathrm{BB}}^H\bm{F}_{\mathrm{RF}}^H\bm{H}^H$}.
	With given $\bm{Q}$ and $\bm{F}_{\mathrm{RF}}$, the optimal $\bm{W}$ for maximizing $\xi(\bm{W},\bm{Q},\bm{F}_{\mathrm{RF}})$ is given by $\bm{W}^{\star}=\bm{E}^{-1}$. Note that (P4) can be shown to be equivalent to itself with $\bm{Q}$ and $\bm{W}$ added as optimization variables and (14b) replaced by $\xi(\bm{W},\bm{Q},\bm{F}_{\mathrm{RF}})\geq \bar{R}$. This equivalent problem can be dealt with using an AO algorithm by iteratively optimizing $\bm{F}_{\mathrm{RF}}$, $\bm{Q}$, and $\bm{W}$, where the only remaining task is to optimize $\bm{F}_{\mathrm{RF}}$ with given $\bm{Q}$ and $\bm{W}$. By substituting $\bm{E}$ into (\ref{reformulate_rate constraint}), we have
	\begin{align}
		\hspace{-3mm}
		\xi(\bm{W}, \bm{Q}, \bm{F}_{\mathrm{RF}})&\!=\!
		\eta-\mathrm{tr}\big(\bm{F}_{\mathrm{BB}}^H\bm{F}_{\mathrm{RF}}^H\bm{B}_1\bm{F}_{\mathrm{RF}}\bm{F}_{\mathrm{BB}}\big)\nonumber\\
		&+\mathrm{tr}\big(\bm{B}_2\bm{F}_{\mathrm{RF}}\big)+\mathrm{tr}\big(\bm{B}_2^H\bm{F}_{\mathrm{RF}}^H\big),
	\end{align}
	where $\bm{B}_1=\bm{H}^H\bm{Q}\bm{W}\bm{Q}^H\bm{H}$, $\bm{B}_2=\bm{F}_{\mathrm{BB}}\bm{W}\bm{Q}^H\bm{H}$, and \textcolor{black}{ $\eta=\log_2|\bm{W}|+N_{\mathrm{S}}-\mathrm{tr}\big(\bm{W}\big)-\sigma_{\mathrm{C}}^2\mathrm{tr}\big(\bm{W}\bm{Q}^H\bm{Q}\big)$}. By applying matrix transformation $\mathrm{tr}(\bm{A}\bm{B}\bm{C}\bm{D})=(\mathrm{vec}(\bm{D}^T))^T(\bm{C}^T\otimes\bm{A})\mathrm{vec}(\bm{B})$, this problem is equivalently expressed as
	\begin{subequations}\label{P4I}
		\begin{align}
			\mbox{(P4-I)}\quad \underset{\bm{v}}{\max}\quad& 
			\bm{v}^H(\bm{R}_{\mathrm{BB}}^T\otimes \bm{A}_1)\bm{v}\label{P4I_obj}\\
			\mbox{s.t.}\quad &\bm{v}^H(\bm{R}_{\mathrm{BB}}^T\otimes \bm{I}_{N_{\mathrm{T}}})\bm{v}\leq P \label{P4I_power constraint} \\   
			&-\bm{v}^H(\bm{R}_{\mathrm{BB}}^T\otimes \bm{B}_{1})\bm{v}+\bm{c}^T\bm{v} \notag\\
			&+ (\bm{c}^T\bm{v})^*\geq \Bar{R}-\eta \label{P4I_rate constraint}\\
			& \bm{v}^H \bm{\Lambda}_m\bm{v}= 1, \quad \forall m,
		\end{align}
	\end{subequations}
	where $\bm{v}=\mathrm{vec}(\bm{F}_{\mathrm{RF}})$, $\bm{c}=\mathrm{vec}(\bm{B}_2^T)$, $\bm{\Lambda}_m\in \mathbb{R}^{N_{\mathrm{T}}N_{\mathrm{RF}}\times N_{\mathrm{T}}N_{\mathrm{RF}} }$ is a matrix whose $m$-th diagnoal element is $1$ and the remaining elements are $0$. 
	
	Problem (P4-I) is still a non-convex quadratically constrained quadratic program (QCQP), since $\bm{R}_{\mathrm{BB}}^T\otimes \bm{A}_1$ is positive semi-definite and $\bm{\Lambda}_m \neq \bm{0}$. To tackle this problem, we propose an FPP-SCA algorithm \cite{mehanna2014feasible}. Specifically, we replace the functions in (P4-I) which introduce non-convexity with their first-order Taylor approximations at point $\bm{z}$, and introduce $\epsilon\gg 1$, $r$, $\bm{p}$, and $\bm{w}$ as penalty variable or slack variables under the FPP-SCA framework. (P4-I) is then approximated as the following convex problem:
	\begin{subequations}
		\begin{align}
			\hspace{-1.5mm}
			\mbox{(P4-II)} \underset{\bm{v}, t, r\geq 0, \bm{p}, \bm{w}}{\max}& 
			t-\epsilon r-\epsilon \|\bm{p}\|_1-\epsilon\|\bm{w}\|_1 \label{P4II_obj}\\
			\mbox{s.t.}\quad & 2\mathfrak{Re}\{\bm{z}^H(\bm{R}_{\mathrm{BB}}^T\otimes \bm{A}_1)\bm{v} \}\!-\!\bm{z}^H(\bm{R}_{\mathrm{BB}}^T\otimes \bm{A}_1)\bm{z}\notag\\
			&\qquad\qquad\geq t-r\\
			&\bm{v}^H(\bm{R}_{\mathrm{BB}}^T\otimes \bm{I}_{N_{\mathrm{T}}})\bm{v}\leq P \label{P4II_power constraint}\\
			&  \bm{v}^H(\bm{R}_{\mathrm{BB}}^T\otimes \bm{B}_{1})\bm{v}-\bm{c}^T\bm{v} - (\bm{c}^T\bm{v})^*\leq -\Bar{R}\notag\\
			&\qquad\qquad +\eta \label{P4II_rate constraint}\\
			& \bm{v}^H \bm{\Lambda}_m\bm{v}\leq 1+p_m, \, p_m\geq0,\forall m\\
			& 2\mathfrak{Re}\big\{\bm{z}^H\bm{\Lambda}_m\bm{v}\big\}-\bm{z}^H\bm{\Lambda}_m\bm{z}\geq 1-w_m, \notag\\
			&\qquad w_m\geq 0,\quad \forall m.
		\end{align}
	\end{subequations}
	Note that the convex problem (P4-II) can be efficiently solved via CVX. Thus, with $\bm{z}$ iteratively updated as the optimal $\bm{v}$ obtained by solving (P4-II) in the previous iteration, a Karush-Kuhn-Tucker (KKT) point to (P4-I) can be obtained as long as the slack variables are converged to zero \cite{mehanna2014feasible}. 
	
	By iteratively obtaining the optimal $\bm{R}_{\mathrm{BB}}$ via solving (P3), the optimal $\bm{Q}$ and $\bm{W}$ via their closed-form expressions, and a high-quality solution of $\bm{F}_{\mathrm{RF}}$ via FPP-SCA, a high-quality suboptimal solution to (P2) can be obtained. Algorithm 1 summarizes this algorithm, for which monotonic convergence is guaranteed.
	
	\vspace{-2mm}
	\begin{algorithm}\label{Alt-Algorithm}
		\LinesNumbered
		\caption{Proposed AO Algorithm for (P2)}
		\KwIn{$\bm{A}_1$, $\bm{H}$, $N_{\mathrm{T}}$, $N_{\mathrm{RF}}$, $N_{\mathrm{U}}$, $N_{\mathrm{S}}$, $P$, \textcolor{black}{$\sigma_{\mathrm{C}}^2$, $\sigma_{\mathrm{S}}^2$}, $\Bar{R}$}
		\KwOut{$\bm{R}_{\mathrm{BB}}$ and $\bm{F}_{\mathrm{RF}}$}
		Randomly initialize $\bm{F}_{\mathrm{RF}}^{(0)}$ with $[\bm{F}_{\mathrm{RF}}]_{m,n}=e^{j\phi_{m,n}}$ \;
		Obtain the optimal $\bm{R}_{\mathrm{BB}}^{(0)}$ based on \cite{xu2023mimo}\;
		Set $t=0$\;
		\Repeat{convergence}
		{
			Given $\bm{F}_{\mathrm{RF}}^{(t)}$ and $\bm{R}_{\mathrm{BB}}^{(t)}$, obtain the optimal $\bm{Q}^{(t)}$\;
			Given $\bm{F}_{\mathrm{RF}}^{(t)}$, $\bm{R}_{\mathrm{BB}}^{(t)}$ and $\bm{Q}^{(t)}$ , obtain the optimal $\bm{W}^{(t)}$\;
			Set $j=0$, initialize $\bm{z}^{(0)}=\mathrm{vec}(\bm{F}_{\mathrm{RF}}^{(t)})$\;
			\Repeat{convergence}
			{
				Set $j=j+1$\;
				Obtain the optimal $\bm{v}^{(j)}$ to (P4-II) via CVX\;
				Set $\bm{z}^{(j)}=\bm{v}^{(j)}$\;
				
			}
			Set $t=t+1$\;
			Set $\bm{F}_{\mathrm{RF}}^{(t)}=\mathrm{reshape}(\bm{v}^{\star})_{N_{\mathrm{T}}\times N_{\mathrm{RF}}}$\;
			Obtain the optimal $\bm{R}_{\mathrm{BB}}^{(t)}$ to (P3) with given $\bm{F}_{\mathrm{RF}}^{(t)}$ based on \cite{xu2023mimo}\;
		}
	\end{algorithm}
	\vspace{-5mm}
	\subsection{Complexity Analysis}
	Finally, we analyze the complexity of Algorithm 1. The complexity of calculating the  matrices $\bm{Q}$ and $\bm{W}$ is $\mathcal{O}(N_{\mathrm{U}}^3)$ and $\mathcal{O}(N_{\mathrm{S}}^3)$, respectively. The complexity of designing the digital beamformer by solving (P3) is $\mathcal{O}(N_{\mathrm{LD}}(N_{\mathrm{RF}}^3+N_{\mathrm{U}}N_{\mathrm{RF}}\min(N_{\mathrm{U}},N_{\mathrm{RF}})))$ where $N_{\mathrm{LD}}$ is the total number of iterations in the Lagrange duality method proposed in \cite{xu2023mimo}. Let $N_{\mathrm{out}}$ denote the number of outer iterations and $N_{\mathrm{in}}$ denote the number of iterations within the FPP-SCA algorithm. The complexity for obtaining the optimal solution to (P2) via Algorithm \ref{Alt-Algorithm} can be shown to be $\mathcal{O}(N_{\mathrm{out}}(N_{\mathrm{U}}^3+N_{\mathrm{LD}}(N_{\mathrm{RF}}^3+N_{\mathrm{U}}N_{\mathrm{RF}}\min(N_{\mathrm{U}},N_{\mathrm{RF}}))+N_{\mathrm{in}}(N_{\mathrm{T}}N_{\mathrm{RF}})^{3.5}))$. 
	\section{Numerical Results}
	In this section, we provide numerical results to evaluate the performance of the proposed hybrid beamforming design. We set $N_{\mathrm{T}}=12$, $N_{\mathrm{R}}=14$, $N_{\mathrm{U}}=8$, and \textcolor{black}{ $\sigma_{\mathrm{C}}^2=\sigma_{\mathrm{S}}^2=-90$} dBm. We further set $P=30$ dBm and \textcolor{black}{ $\frac{P|\alpha|^2L}{\sigma_{\mathrm{S}}^2}=-5$} dB. The BS-target distance is set as $r=40$ m. We set $\bar{R}=5$ bps/Hz and $N_{\mathrm{RF}}=3$ unless specified otherwise. In practical scenarios, the angle distribution of a target is usually concentrated around one or multiple nominal angles. Therefore, we assume that the PDF of $\theta$ follows a Gaussian mixture model which is the weighted summation of $K\geq 1$ Gaussian PDFs and is given by $
		p_{\Theta}(\theta)=\sum_{k=1}^K\frac{p_k}{\sqrt{2\pi}\sigma_k}e^{-\frac{(\theta-\theta_k)^2}{2\sigma_k^2}}$. Specifically, $\theta_k\in[-\frac{\pi}{2}, \frac{\pi}{2})$ and $\sigma_k^2$ are the mean and variance of the $k$-th Gaussian PDF, respectively; $p_k$ denotes the weight that satisfies $\sum_{k=1}^Kp_k=1$. We set $K=5$; $\theta_1=-0.81$, $\theta_2=-0.72$, $\theta_3=-0.18$, $\theta_4=0.75$, $\theta_5=0.93$; $\sigma_1^2=10^{-3}$, $\sigma_2^2=10^{-2.5}$, $\sigma_3^2=10^{-3}$, $\sigma_4^2=10^{-2.5}$, $\sigma_5^2=10^{-3}$; $p_1=0.30$, $p_2=0.20$, $p_3=0.11$, $p_4=0.18$, $p_5=0.21$. 
	
	For communication, we consider a Rician fading channel modeled as \textcolor{black}{ $\bm{H}=\sqrt{\beta_{\mathrm{C}}/(K_{\mathrm{C}}+1)}(\sqrt{K_{\mathrm{C}}}\bm{H}_{\mathrm{LoS}}+\bm{H}_{\mathrm{NLoS}})$}, where \textcolor{black}{$\beta_{\mathrm{C}}=\frac{\beta_0}{r_{\mathrm{U}}^{3.5}}$} denotes the average channel power with $\beta_0=-30$ dB and $r_{\mathrm{U}}=400$ m denoting the BS-user distance; \textcolor{black}{$K_{\mathrm{C}}=-8$} dB denotes the Rician factor. The LoS component is set as $\bm{H}_{\mathrm{LoS}}=\bm{b}_{\mathrm{U}}(\theta_{\mathrm{U}})\bm{a}(\theta_{\mathrm{U}})^H$, where $\bm{b}_{\mathrm{U}}(\theta_{\mathrm{U}})$ denotes the steering vector of the user receive antenna array under the ULA layout, and $\theta_{\mathrm{U}}\!=\!0.36$ denotes the user's angle. Moreover, $[\bm{H}_{\mathrm{NLoS}}]_{m,n}\!\sim\! \mathcal{CN}(0,1)$, $\forall m, n$. Due to the limited space, we focus on evaluating the performance of Algorithm 1.
	
	\begin{figure}[t]
		\centering
		\includegraphics[width=0.4\textwidth]{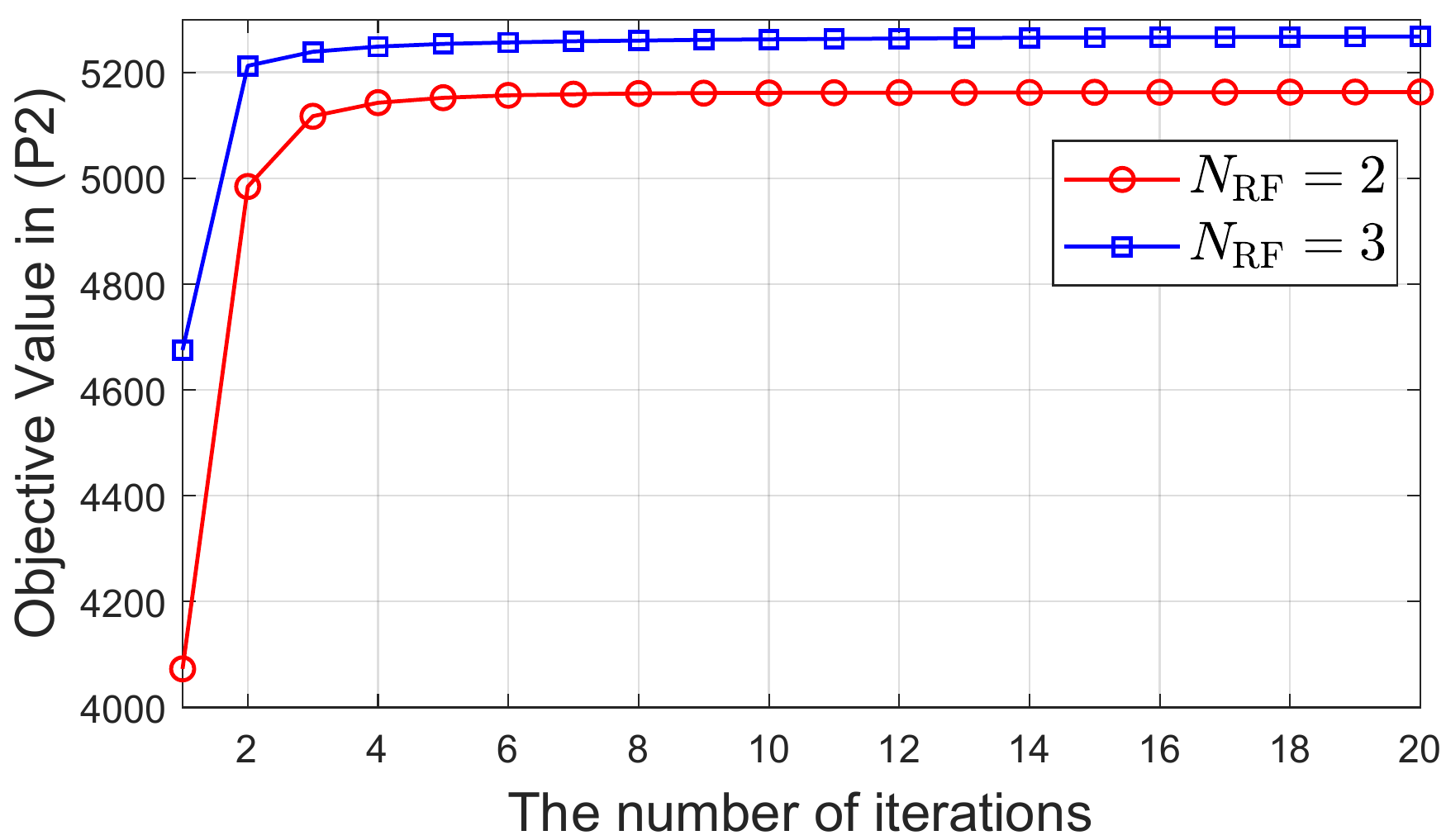}
		\vspace{-3mm}
		\caption{Convergence behavior of Algorithm 1.}
		\label{convergence_behavior}
		\vspace{-2mm}
	\end{figure}
	
\begin{figure}[t]
		\vspace{-1.8mm}
		\centering
		\includegraphics[width=0.38\textwidth]{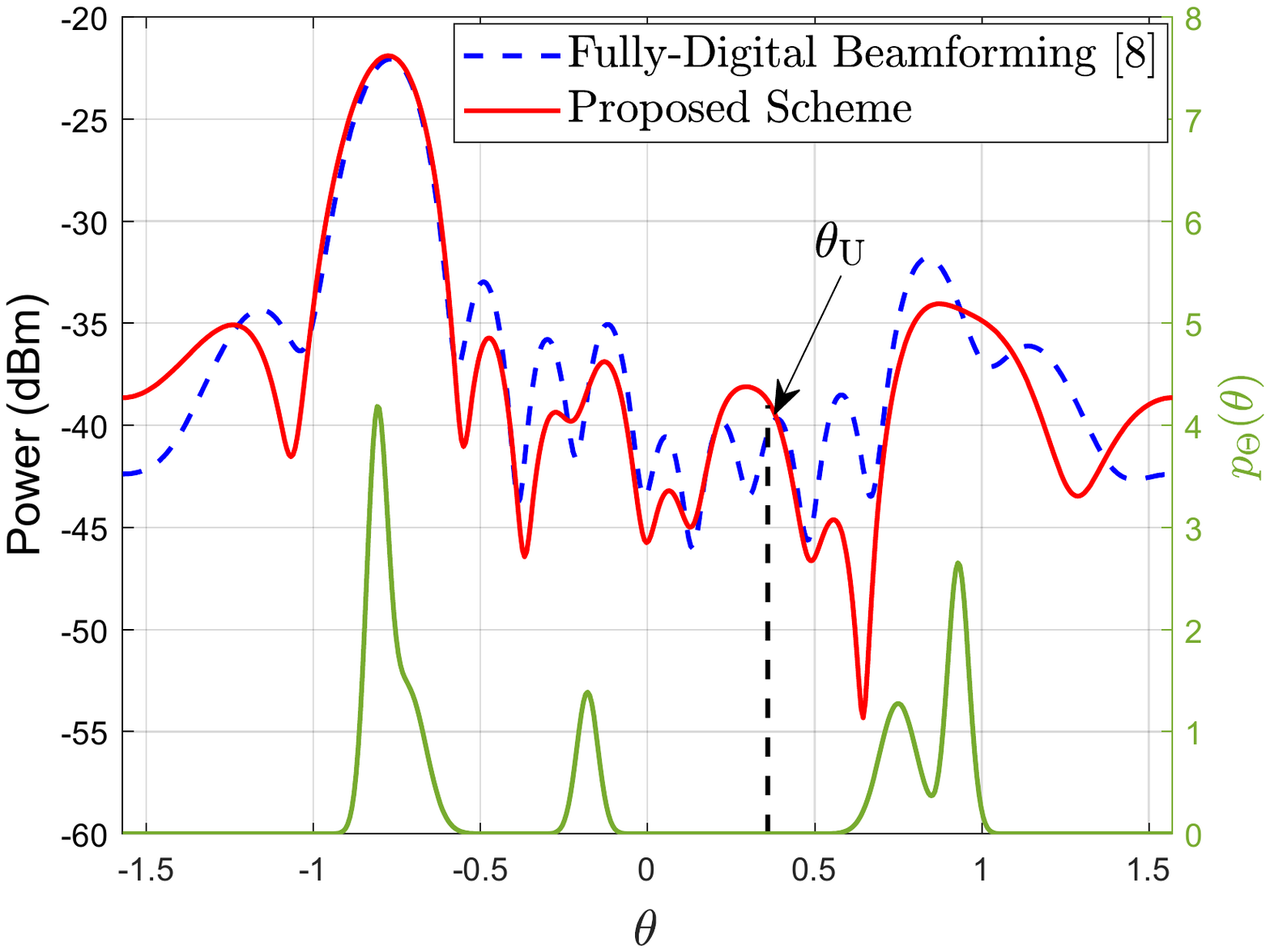}
		\vspace{-4.5mm}
		\caption{Radiated power pattern and $p_{\Theta}(\theta)$ over different angles.}
		\label{Radiatedpowerpattern}
		\vspace{-4.5mm}
	\end{figure}

	First, we evaluate the convergence behavior of Algorithm 1 in Fig. \ref{convergence_behavior}. It is observed that Algorithm 1 converges monotonically and rapidly for both values of $N_{\mathrm{RF}}$'s. Then, Fig. \ref{Radiatedpowerpattern} shows the radiated power pattern and $p_{\Theta}(\theta)$ over different angles. It is observed that the radiated power pattern achieved by both the optimal fully-digital beamforming \cite{xu2023mimo} and the proposed hybrid beamforming can focus the power over the target's possible angles with high probability densities as well as the the user's angle, which demonstrates the effectiveness of our proposed AO algorithm.
	
	\begin{figure}[t]
		\centering
		\includegraphics[width=0.38\textwidth]{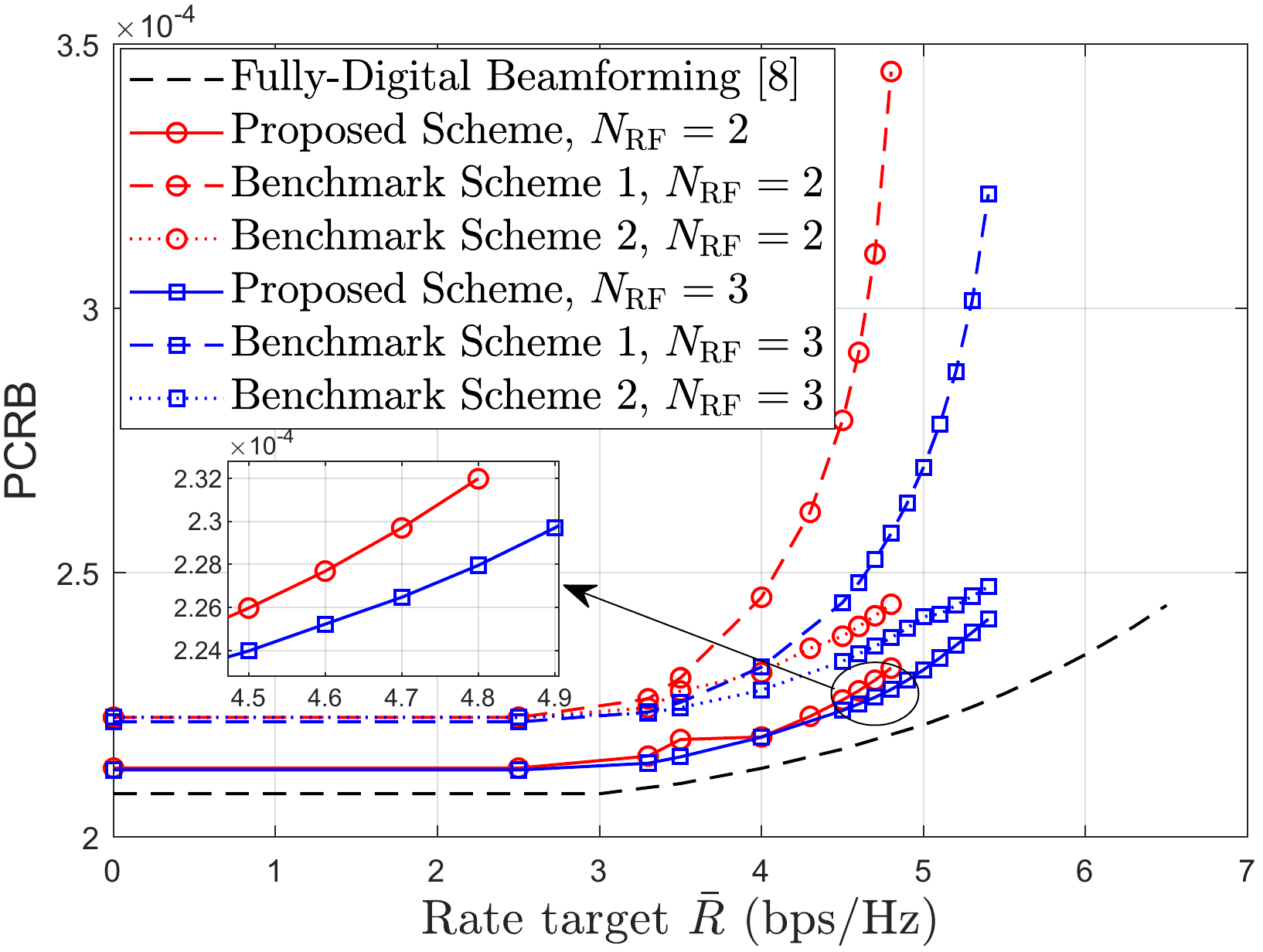}
		\vspace{-4mm}
		\caption{$\mathrm{PCRB}$ versus the communication rate target.}
		\label{PCRB_fig}
		\vspace{-6.5mm}
	\end{figure}    
	
	In Fig. \ref{PCRB_fig}, we show the PCRB versus the communication rate target $\Bar{R}$ for the proposed scheme, optimal fully-digital beamforming \cite{xu2023mimo}, and the following benchmark schemes:
	\vspace{-1mm}
	\begin{itemize}
		\item {\bf{Benchmark Scheme 1: Heuristic design}.} In this scheme, the analog beamformer is heuristically designed to increase the array gain at the user and the target. Let $\mathbf{f}_{\mathrm{RF},i}$ denotes the $i$-th column of $\bm{F}_{\mathrm{RF}}$ designed as
		\begin{equation}
			\mathbf{f}_{\mathrm{RF},i}=
			\begin{cases}
				\bm{a}(\theta_{\mathrm{U}}), & i=1,\\
				\bm{a}(\theta_{i,\mathrm{max}}), & i=2,\cdots,N_{\mathrm{RF}},
			\end{cases}
		\end{equation}
		where $\theta_{i,\mathrm{max}}$ is the element in $\{\theta_k\}_{k=1}^K$ with the $(i-1)$-th largest $p_k$. $\bm{R}_{\mathrm{BB}}$ is designed by solving (P3).
		\item {\bf{Benchmark Scheme 2: Highest-probability angle based design}.} In this scheme, we design $\bm{F}_{\mathrm{RF}}$ and $\bm{R}_{\mathrm{BB}}$ to maximize the radiated power $\frac{\beta_0}{r^2}\bm{a}^H(\theta_{\mathrm{max}})\bm{F}_{\mathrm{RF}}\bm{R}_{\mathrm{BB}}\bm{F}_{\mathrm{RF}}^H\bm{a}(\theta_{\mathrm{max}})$ corresponding to the angle with highest probability, i.e., $\theta_{\mathrm{max}}\!=\!\arg \max p_{\Theta}(\theta)$.
	\end{itemize}
	\addtolength{\topmargin}{0.02in}
	It is observed that the PCRB increases as the communication rate target becomes larger for all schemes, due to the necessity of allocating the limited power and spatial resources at the BS transmitter between the communication and sensing functions which leads to the non-trivial PCRB-rate trade-off. Moreover, increasing $N_{\mathrm{RF}}$ can improve both the communication and sensing performances. On the other hand, our proposed AO algorithm achieves close performance to the optimal fully-digital beamforming, and significantly outperforms Benchmark Schemes 1 and 2. This proves the effectiveness of our proposed hybrid beamforming design.
	\section{Conclusions}
	This paper studied the hybrid beamforming optimization problem in a MIMO ISAC system, which aims to simultaneously communicate with a multi-antenna user and sense the unknown and random location of a point target by exploiting its prior distribution information. For the sensing-only case, we analytically proved that two RF chains suffice for hybrid beamforming to achieve the same optimal sensing performance as fully-digital beamforming in terms of PCRB. We further developed an efficient algorithm for designing hybrid beamforming with a single RF chain. Furthermore, we formulated the hybrid beamforming optimization problem to minimize a PCRB upper bound subject to a communication rate target. Despite the non-convexity of this problem, we devised an AO algorithm based on the FPP-SCA method to obtain a high-quality suboptimal solution. The effectiveness of the proposed AO algorithm was validated via numerical results.
	
	\bibliographystyle{IEEEtran}
	\bibliography{Reference}	
\end{document}